\documentclass[10pt,aps,prb,twocolumn,superscriptaddress,showpacs]{revtex4-1}
\usepackage{graphicx}
\usepackage{amssymb}
\usepackage{amsmath}
\usepackage{appendix}
\usepackage{comment}

\begin{document}

\title{Local thermomagnonic torques in two-fluid spin dynamics}

\author{Benedetta Flebus}
\affiliation{Institute for Theoretical Physics and Center for Extreme Matter and Emergent Phenomena, Utrecht University, Leuvenlaan 4, 3584 CE Utrecht, The Netherlands}
\author{Pramey Upadhyaya}
\affiliation{Department of Physics and Astronomy, University of California, Los Angeles, California 90095, USA}
\author{Rembert A. Duine}
\affiliation{Institute for Theoretical Physics and Center for Extreme Matter and Emergent Phenomena, Utrecht University, Leuvenlaan 4, 3584 CE Utrecht, The Netherlands}
\affiliation{Department of Applied Physics, Eindhoven University of Technology, PO Box 513,
5600 MB, Eindhoven, The Netherlands}
\author{Yaroslav Tserkovnyak}
\affiliation{Department of Physics and Astronomy, University of California, Los Angeles, California 90095, USA}


\begin{abstract}
We develop a  general phenomenology describing the interplay between coherent and incoherent dynamics in ferromagnetic insulators. Using the Onsager reciprocity and Neumann's principle, we derive expressions for the local thermomagnonic torques exerted by thermal magnons on the order-parameter dynamics and the reciprocal pumping processes, which are in close analogy to the spin-transfer torque and spin pumping at metallic interfaces. Our formalism is applicable to general long-wavelength dynamics and, although here we explicitly focus on ferromagnetic insulators possessing \textit{U}(1) symmetry, our approach can be easily extended to other classes of magnetic materials. As an illustrative example, we apply our theory to investigate a domain wall floating over a spin superfluid, whose dynamics is triggered thermally at the system's edge. Our results demonstrate that the local pumping of coherent spin dynamics by a thermal magnon gas offers an alternative route - with no need for conducting components and thus devoid of Ohmic losses - for the control and manipulation of topological solitons. 
\end{abstract}

\maketitle

\section{Introduction}

The interaction between spin-polarized electron transport and magnetization dynamics via spin-transfer torques\cite{slonczewskiJMMM96,*bergerPRB96} and spin pumping\cite{tserkovPRL02sp} has been investigated for almost two decades now. It paved the way for the manipulation of magnetization textures and dynamics without the deployment of external magnetic fields.\cite{brataasCHA12}  Recently, much enthusiasm has been bolstered by the possibility of attaining similar outcomes by means of thermal control. Thermally-driven magnetization dynamics could be achieved through laser pulsing, as well as through heat diffusion, thereby removing the need for an electronic medium altogether.\cite{bauerNATM12}

In magnetic insulators, a thermal bias triggers a pileup of thermal magnons via the spin Seebeck effect.\cite{uchidaAPL10,*xiaoPRB10,*adachiPRB11} This incoherent magnon cloud can relax by transferring spin angular momentum to the magnetic order parameter and thus resulting in a local (thermomagnonic) spin-transfer torque.\cite{benderPRB16,tserkovPRB16} The latter may then launch nonequilibrium spin textures, opening up new prospects for thermally-driven nonvolatile magnetic memories and logic with potentially little net dissipation.

In this work, via Neumann's principle and the Onsager reciprocity relations,  we develop a general formalism describing the local thermomagnonic torques exerted on the magnetic order parameter and the backaction of the coherent dynamics on the thermal magnons. These reciprocal phenomena are in close analogy to the spin-transfer torque and spin pumping in metallic multilayers.\cite{tserkovRMP05} Our phenomenology is suited to describe the interplay between diffusive and collective (Landau-Lifshitz--type) dynamics for general spin textures, providing a generalization of previous results.\cite{benderPRB16,tserkovPRB16}

To simplify our discussion, we focus on the simplest nontrivial case yielding local thermomagnonic torques, i.e., axially-symmetric (either easy- or hard-axis) magnetic systems. The hard-axis case has been proposed\cite{soninJETP78,*soninAP10} for hosting a spin superfluid, which is rooted in the Goldstone mode associated with the 
spontaneous \textit{U}(1) symmetry breaking.\cite{takeiPRL16,flebusPRL16} 
In contrast to the exponentially-decaying flow of thermal magnons, the spin superfluid can transmit spin transport over long distances. This has been exploited recently by Upadhyaya \textit{et al.},\cite{upadhyayaCM16} who suggested that a hard-axis magnet can efficiently transfer spin angular momentum between a metallic spin reservoir and a distant domain wall. Here, we employ our phenomenology to extend their proposal to a domain wall  driven by a thermally-activated superfluid dynamics.

\section{Local thermomagnonic torques}

In this section, we construct a general  phenomenology describing the coupling between  magnetic order-parameter dynamics and a quasi-equilibrium cloud of thermal magnons. 
Specifically, we consider a magnetic insulator, whose spin density is given in the ground state by $\mathbf{s}=s \mathbf{n}$, $s$ being the saturated spin density (in units of $\hbar$) and $\mathbf{n}$ the spin-density orientation. Finite temperature gives rise to the fluctuations $\delta\hat{\mathbf{s}}=\hat{\mathbf{s}}-\langle \hat{\mathbf{s}} \rangle$, where $\hat{\mathbf{s}}$ is the spin-density operator. These are composed of thermal magnons, whose density $\tilde{n}$ reduces the magnitude of the spin density to $\tilde{s}\equiv s-\tilde{n}$. Here, we will assume that the interactions within the thermal magnon cloud are fast enough (compared to the pumping and relaxation processes) to equilibrate them to a common temperature $T$ and chemical potential $\mu$. We are supposing the temperature to be large compared to the anisotropy fields, to that the magnons are of the exchange type and carry spin $-\hbar$ along $\mathbf{n}$.

If the coherent texture is smooth on the scale of the thermal-magnon wavelength, the hydrodynamic variables that describe the system are the orientation $\mathbf{n}$ of the order parameter and the thermal-magnon density $\tilde{n}$, which together parametrize the total (three-component) spin density $\langle \hat{\mathbf{s}} \rangle\equiv (s-\tilde{n}) \mathbf{n}$. In terms of these variables, the instantaneous state of the magnet can be described by a free-energy functional $\mathcal{F}[\mathbf{n},\tilde{n}]$. The effective (Landau-Lifshitz) transverse field $\mathbf{H}\equiv \delta_{\mathbf{n}} \mathcal{F}[\mathbf{n}, \tilde{n}]$ and the chemical potential $\mu\equiv\delta_{\tilde{n}} \mathcal{F}[\mathbf{n}, \tilde{n}]$  are the  forces conjugate to the variables $\mathbf{n}$ and $\tilde{n}$, respectively.

Within the linear response, the relations between the rates $\dot{\mathbf{n}}$ and $\dot{\tilde{n}}$  and the forces can be written as
\begin{align}
\begin{pmatrix} \dot{\mathbf{n}} \\ \dot{\tilde{n}} \end{pmatrix}= \begin{pmatrix}
\mathbf{L}^{\mathbf{n} \mathbf{n}} & \mathbf{L}^{\mathbf{n} \tilde{n}} \\ \mathbf{L}^{\tilde{n} \mathbf{n}} &  \mathbf{L}^{\tilde{n} \tilde{n}} 
\end{pmatrix} \begin{pmatrix}
\mathbf{H} \\ \mu
\end{pmatrix} \equiv \mathbf{L}  \begin{pmatrix}
\mathbf{H} \\ \mu
\end{pmatrix} \,,
\label{eq2}
\end{align}
where we have introduced  the $3\times3$ linear-response matrix $\mathbf{L}$, per each point in space. ($\mathbf{L}^{\mathbf{n} \mathbf{n}}$ is a $2\times2$ block etc.) Leaving the relaxation processes aside for the moment, the decoupled orientational dynamics obey the Landau-Lifshitz equation:\cite{landau1935} 
\begin{align}
\hbar \dot{\mathbf{n}}=\frac{1}{\tilde{s} }\mathbf{H} \times  \mathbf{n} \,.
\label{equation1}
\end{align}
The decoupled dynamics of the incoherent magnon cloud is treated diffusively:
\begin{align}
\dot{\tilde{n}}=-\boldsymbol{\nabla}\cdot\tilde{\mathbf{j}}\,,
\label{equation2}
\end{align}
where we have defined (in the absence of thermal gradients, for now) $\tilde{\mathbf{j}}= - \sigma \boldsymbol{\nabla} \mu $ as the magnon flux, with $\sigma$ being the magnon conductivity.\cite{flebusPRL16, ludo2016}  
The kinetic (matrix-valued) coefficients $\textbf{L}^{\textbf{n} \textbf{n}}$ and $\mathbf{L}^{\tilde{n} \tilde{n}}$ can be easily read off from Eqs.~(\ref{equation1}) and (\ref{equation2}). The off-diagonal coefficient $\mathbf{L}^{\mathbf{n} \tilde{n}}$ describes the thermomagnonic torque exerted by the thermal magnons on the orientational dynamics. Its reciprocal counterpart is the pumping of the magnon gas by the coherent magnetic precession, which is described by the coefficient $\mathbf{L}^{\tilde{n} \mathbf{n}}$.

The off-diagonal linear-response coefficients are connected via the Onsager reciprocity, which dictates that\cite{onsagerPR31p1,*onsagerPR31p2}
\begin{align}
 [\mathbf{L}^{\mathbf{n} \tilde{n}} (\mathbf{n})]_{ij}=-[\mathbf{L}^{\tilde{n} \mathbf{n}}(-\mathbf{n} )]_{ji}\,,
 \label{Onsag1}
 \end{align}
where the minus signs stems from different time-reversal transformations of $\mathbf{n}$ and $\tilde{n}$.
Let us next write the equation of
motion for $\mathbf{n}$ due to thermomagnonic torques as
\begin{align}
\hbar \dot{\mathbf{n}}=\mathbf{n}  \times \mathbf{h}(\mu, \mathbf{n}, \dot{\mathbf{n}})\,,
\label{Equaz6}
\end{align}
where $\mathbf{h}(\mu, \mathbf{n}, \dot{\mathbf{n}})\perp\mathbf{n}$ is a linear function of the nonequilibrium arguments  $\mu$ and $\dot{\mathbf{n}}$. Terms $\propto \dot{\mathbf{n}}$ in Eq.~(\ref{Equaz6}) contribute to the coefficient $\mathbf{L}^{\mathbf{n} \mathbf{n}}$. Their form is restricted by the Onsager reciprocal relations between the components of the transverse magnetization dynamics, i.e.,
\begin{align}
\mathbf{L}^{\mathbf{n} \mathbf{n}}(\mathbf{n})=[\mathbf{L}^{\mathbf{n} \mathbf{n}} (-\mathbf{n})]^{\text{T}} \,,
\label{Onsager2}
\end{align}
where $^{\text{T}}$ denotes matrix transpose. In addition to the requirements imposed by the reciprocity relations (\ref{Onsag1}) and (\ref{Equaz6}), the form of $\mathbf{h}(\mu, \mathbf{n}, \dot{\mathbf{n}})$ must be constrained by the structural symmetries of the system.\cite{birssBOOK66}

In the following, we restrict our attention to insulating magnets which retain \textit{U}(1) symmetry, typical examples of which are the simple easy-plane and easy-axis ferromagnets. In these systems, due to the  rotational invariance around the $z$ axis,  Neumann's principle requires that 
\begin{align}
\mathbf{h}(\mathcal{R}_{z} \mathbf{n}, \mathcal{R}_{z} \dot{\mathbf{n}})=\mathcal{R}_{z} \{ \mathbf{h}(\mathbf{n},\dot{\mathbf{n}} )\}\,,
\label{neumann}
\end{align}
where $\mathcal{R}_{z}(\theta)$ is the SO(3) rotation matrix by angle $\theta$ around the $z$ axis. The \textit{U}(1) symmetry, furthermore, enforces the conservation of the $z$ component of the total angular momentum associated with the total spin density, i.e.,
\begin{align}
\tilde{s}\dot{n}_z - \dot{\tilde{n}} n_{z}=0 \,,
\label{spinconserv}
\end{align}
where $n_{z}\equiv\hat{\mathbf{z}}\cdot\mathbf{n}$. Imposing the constrains (\ref{Onsag1})-(\ref{spinconserv}) finally yields, after straightforward manipulations:
\begin{align}
\hbar \dot{\mathbf{n}} =& \eta' n_z (\hbar n_{z} \dot{\mathbf{n}}-\mu \mathbf{n}\times\hat{\mathbf{z}})  - \eta n_z \mathbf{n} \times (\hbar n_{z} \dot{\mathbf{n}}-\mu \mathbf{n}\times\hat{\mathbf{z}}) \nonumber \\
&+\frac{1}{\tilde{s}}\mathbf{H} \times \mathbf{n} \,, \label{equaz9} \\
\dot{\tilde{n}} =& \eta' \tilde{s} n_z\dot{n}_z - \eta\frac{\tilde{s}}{\hbar}\hat{\mathbf{z}} \cdot\mathbf{n}\times(\hbar n_{z} \dot{\mathbf{n}} - \mu \mathbf{n} \times \hat{\mathbf{z}}) -\boldsymbol{\nabla} \cdot \tilde{\mathbf{j}}\,.
\label{equaz10}
\end{align}
where $\eta$ and $\eta'$ are some even function of $n_{z}$. Since we are working at linear response, $\tilde{s}$ here can be taken to be the equilibrium spin density at the ambient temperature $T$.
 
We next proceed to restore the relaxation mechanisms, both for the precessional dynamics and the magnon density. Microscopically, these are rooted in the relativistic corrections, such as spin-orbit coupling, which would affect the conservation of the $z$ component of the spin angular momentum. We thus relax the constraint \eqref{spinconserv} when including the relaxation terms, while not revising our derivation of Eqs.~(\ref{equaz9}) and (\ref{equaz10}). The underlying premise of such an approach is that the relaxation processes are usually weak enough that we can start by disregarding their role in the spin transfer between the coherent and incoherent dynamics.

The damping terms naturally appear in the Gilbert and Bloch forms for $\mathbf{n}$ and $\tilde{n}$, respectively, which append Eqs.~\eqref{equaz9} and \eqref{equaz10} as follows:
\begin{align}
\hbar \dot{\mathbf{n}} =& \eta' n_z (\hbar n_{z} \dot{\mathbf{n}}-\mu \mathbf{n}\times\hat{\mathbf{z}})  - \eta n_z \mathbf{n} \times (\hbar n_{z} \dot{\mathbf{n}}-\mu \mathbf{n}\times\hat{\mathbf{z}}) \nonumber \\
&+\frac{1}{\tilde{s}}\mathbf{H} \times \mathbf{n} - \alpha \hbar \mathbf{n} \times \dot{\mathbf{n}}\,, \label{EQ12} \\
\dot{\tilde{n}} =& \eta'\tilde{s}  n_z\dot{n}_z - \eta\frac{\tilde{s}}{\hbar}\hat{\mathbf{z}} \cdot\mathbf{n}\times(\hbar n_{z} \dot{\mathbf{n}} - \mu \mathbf{n} \times \hat{\mathbf{z}}) -\boldsymbol{\nabla} \cdot \tilde{\mathbf{j}} \nonumber \\
& - \gamma \mu\,.
\label{EQ13}
\end{align}
Here, $\alpha$ and $\gamma$ parametrize the Gilbert damping and the ($T_1$) Bloch relaxation of magnons, respectively. While $\alpha$ and $\gamma$ can generally depend on $n^{2}_{z}$ (and have a tensorial form, according to the axial symmetry), we will for simplicity be considering the limit when they are mere constants. Note that the general  thermomagnonic torques $\propto\eta$ in Eqs.~(\ref{EQ12}) and (\ref{EQ13}) reproduce the results of Ref.~[\onlinecite{tserkovPRB16}] for $n_{z}\approx-1$ (considered there).\cite{Note1} The terms $\propto \eta'$ on the right-hand side of both equations have been omitted in Ref.~[\onlinecite{tserkovPRB16}], which we will similarly do hereafter. Indeed, in Eq.~(\ref{EQ12}), the term $\propto \eta' \dot{\mathbf{n}}$ can be combined with the left-hand side, merely leading to a small rescaling of the equation if (which is natural to expect) $\eta' \ll 1$, while the term $\propto \eta' \mu$ gives rise to a field-like torque, which does not play a substantial role in the dynamics that we are interested in. The term $\propto \eta' \dot{n}_{z}$ in Eq.~(\ref{EQ13}) is inoperative in a steady state with $n_z={\rm const}$, which is the case we will be focusing on. As a final simplification, we will take the remaining coefficient $\eta$ (which microscopically stems from the axial anisotropy\cite{benderPRB16}) to be a constant, in the spirit of our treatment of $\alpha$ and $\gamma$. 
 
\section{Domain wall floating on a superfluid}

Let us now turn to a concrete application of the formalism derived 
in the previous section. Specifically, we will investigate the coupling 
between a domain wall and a spin superfluid. Our setup is similar to that of Ref.~[\onlinecite{upadhyayaCM16}], except that the superfluid dynamics are here triggered thermally. It is accomplished by a thermomagnonic torque exerted by a pileup of thermal magnons, which is induced by a local heat source.

The key ingredient for the realization of a system supporting both zero modes, the spin superfluid and the domain wall, is the spontaneous breaking of the \textit{U}(1)$\times Z_{2}$ composite symmetry, with \textit{U}(1) standing for the rotations around the $z$ axis (which would define a spin superfluid within, e.g., an easy-plane magnet) and $Z_2$ for the time reversal (which would govern domain walls within, e.g.,  an easy-axis magnet). A weakly exchange coupled bilayer of an easy-plane and an easy-axis magnetic films proposed in Ref.~[\onlinecite{upadhyayaCM16}] is one such system that could be easily engineered. See Fig.~1 for a schematic.

\begin{figure}[t]
\includegraphics[width=0.9\linewidth]{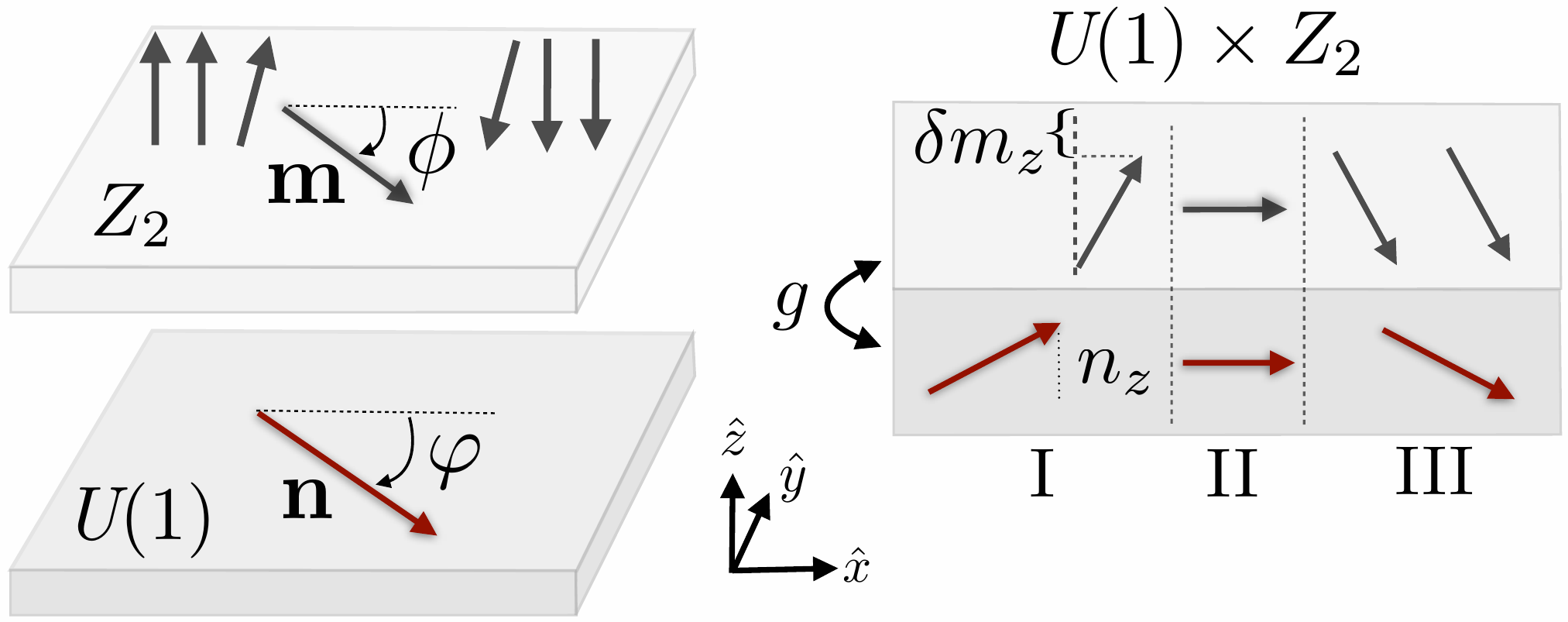}
\caption{An easy-axis magnet exhibits an Ising-like order, with the global ground state oriented either up or down in spin space; a domain wall, resulting from this $Z_{2}$ symmetry breaking, separates the up and down domains (which are related by the time reversal). A superfluid arises in an easy-plane magnet from the spontaneous \textit{U}(1) symmetry breaking. When the layers are coupled together by a weak exchange interaction $\propto g$,  the resulting bilayer displays a composite \textit{U}(1)$ \times Z_{2}$ symmetry breaking. The coupling induces a tilt $n_{z}$ ($\delta m_{z}$) of the order parameter $\mathbf{n}$ ($\mathbf{m}$) in regions I and III, while it locks together the orientations of the order parameters in region II. Here $\varphi$ ($\phi$) is the azimuthal angle of the easy-axis (easy-plane) order parameter $\mathbf{n}$ ($\mathbf{m}$).}
\label{1DDIS}
\end{figure}

While the easy-plane magnet hosts a spin superfluid, the ground state of an easy-axis magnet 
breaks the time-reversal ($Z_{2}$) symmetry, harboring a domain wall 
as a topologically-stable defect. The exchange coupling  $\propto g$ 
between the two layers acts as an effective magnetic field on the 
easy-plane magnet: it tilts the order parameter, $\mathbf{n}$, out-of-plane, resulting in a finite  $n_{z}$. The latter enables the conversion of thermal magnons into coherent spin dynamics via the thermomagnonic torques $\propto\eta\mu$ in Eq.~(\ref{EQ12}). In the domain-wall region, the exchange coupling locks the orientations of the easy-plane and the easy-axis order parameters, allowing for an efficient transfer of angular momentum. This, finally, gives rise to the domain-wall motion, as argued in Ref.~[\onlinecite{upadhyayaCM16}].

\subsection{Model} 

We consider a bilayer  of an  easy-axis ferromagnet (of thickness $\bar{t}$) coupled to an easy-plane ferromagnet (of thickness  $t$), as sketched in Fig.~2(a). Our analysis can also be straightforwardly generalized to an easy-axis ferromagnet$|$easy-plane antiferromagnet heterostructure\cite{upadhyayaCM16} or essentially any \textit{U}(1)$\times Z_2$-breaking system of the type sketched in Fig.~\ref{1DDIS}.

\begin{figure}[t]
\includegraphics[width=0.9\linewidth]{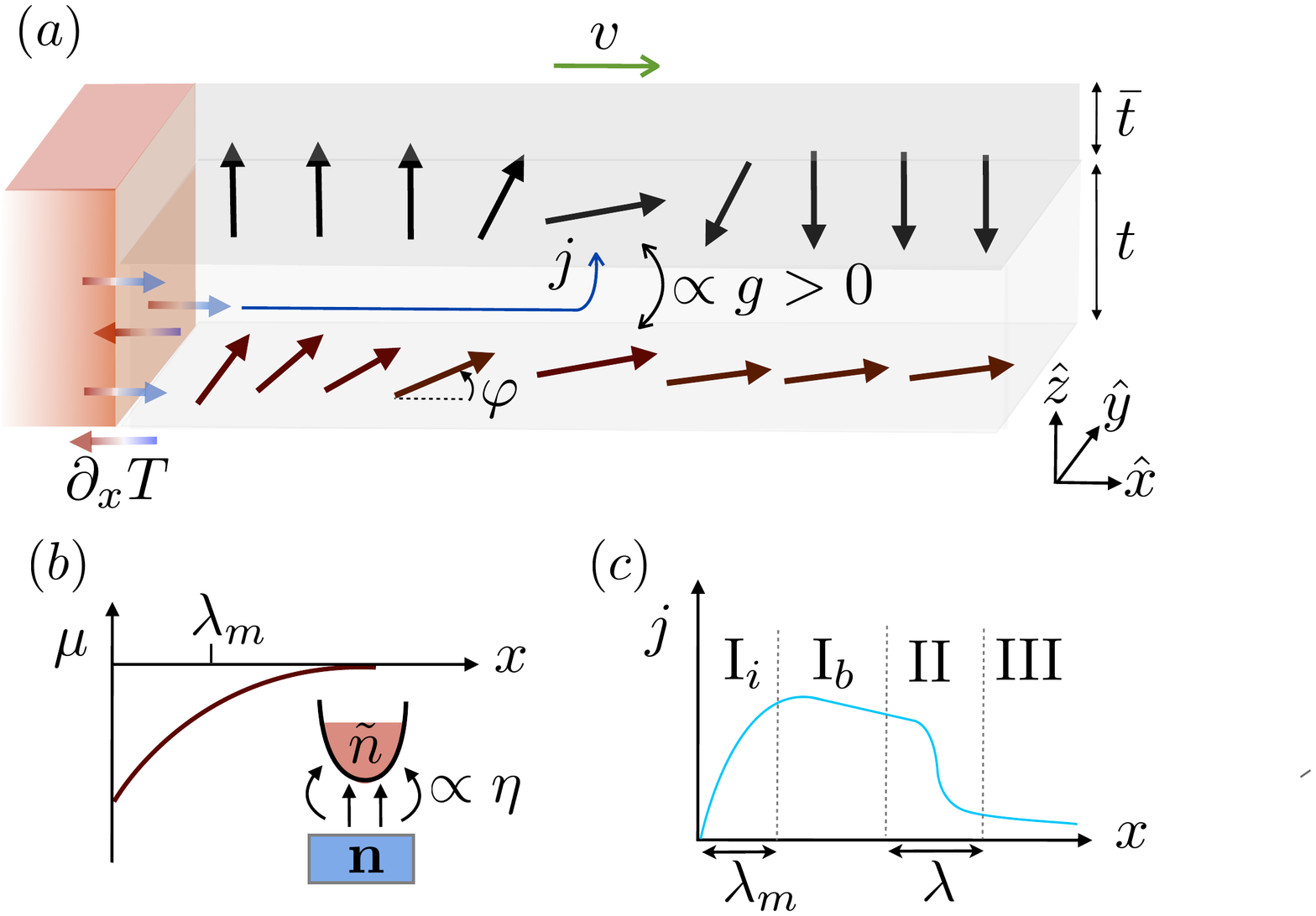}
\caption{(a) A bilayer of an easy-plane ferromagnet of thickness $t$, coupled (with coupling strength $g>0$) to an easy-axis ferromagnet of thickness $\bar{t}$. At the left edge, a heat conductor induces the heat flux $\propto-\partial_{x} T$, which, in turn, activates the superfluid current $j \propto -\partial_{x} \varphi$. The coupling $\propto g$ locks the easy-plane and easy-axis orientations in the domain-wall region, interrupting the superfluid flow. The superfluid current is then absorbed by the domain wall, inducing its motion with velocity $v$. (b) Out-of-equilbrium chemical potential profile decaying away from the left edge with the diffusion length $\lambda_{m}$. The thermal magnon density $\tilde{n}$ equilibrates by exerting a torque $\propto \eta$ on the order parameter $\mathbf{n}$. (c) Superfluid current profile. The current increases and saturates exponentially in the interfacial region $\text{I}_{i}$,  decays linearly in the bulk region $\text{I}_{b}$, is absorbed by the domain wall in region II, with the remaining current decaying linearly within region III.}
\label{figure2}
\end{figure}

A biased heat conductor at the left contact induces an accumulation of thermal magnons in the easy-plane layer, which is localized within the spin-diffusion length $\lambda_{m}$.\cite{flebusPRL16} To simplify our discussion, we suppose that $t \ll \lambda_{m} \ll L$, where $L$ is the magnetic bilayer length in the $x$ direction. The thermally-induced nonequilibrium magnon density $\tilde{n}$ exerts a torque $\propto\eta$ over the spin-diffusion length, triggering superfluid dynamics in the easy-plane magnet. The spin transport is subsequently carried along the $x$ axis by means of coherent precession of $\mathbf{n}$ in the $xy$ plane.\cite{upadhyayaCM16} The thermal magnons hosted in the easy-axis magnet can also exert a torque on the superfluid across the interface. Here, however, in the limit of a weak interlayer coupling, i.e. $|g| \ll K t \; (\bar{K} \bar{t})$ - with $\bar{K}$ ($K$) being magnetic anisotropy of the easy-axis (easy-plane) magnet-, we can neglect it, as it scales as $\propto g^{2}$.

The free-energy density (per unit of area in the $xy$ plane) describing our bilayer is 
\begin{align}
\mathcal{F}[\mathbf{m}, \mathbf{n}, \tilde{n}]= &\bar{A}\bar{t}(\partial_{x} \mathbf{m})^{2}/2 - \bar{K}\bar{t} m^{2}_{z}/2\nonumber\\
&+ At(\partial_{x} \mathbf{n})^{2}/2 + Kt n^{2}_{z}/2\nonumber\\
& + U_{\text{int}}[\mathbf{m},\mathbf{n}] + U[\tilde{n}]\, ,
\label{EQZ1}
\end{align}
where $\bar{A}$ ($A$) is the exchange stiffness  of the easy-axis (easy-plane) magnet,  and we supposed quasi-one-dimensional textures along the $x$ axis. The interfacial exchange interaction $U_{\text{int}}=-g\, \mathbf{m} \cdot \mathbf{n}$ couples the order parameters of the two magnets with the coupling strength $g$. $U[\tilde{n}]$ is the thermal-magnon free energy, taken to be decoupled from the order parameters, as our focus is on the dissipative spin torques. $\mathbf{m}=(\sin\theta \cos \phi, \sin 
\theta \sin \phi, \cos \theta)$ is the unit vector oriented along the 
direction of the spin density in the easy-axis ferromagnet, parametrized by the spherical angles $\theta$ and $\phi$.
The spin-density orientation of the easy-plane ferromagnet $
\mathbf{n}=(\sqrt{1-n^{2}_{z}} \cos\varphi, \sqrt{1-n^{2}_{z}} \sin
\varphi, n_{z})$ is parametrized by the azimuthal angle $
\varphi$ and the $z$ projection $n_{z}$. The chemical potential $\mu$ is contained in the dependence $U[\tilde{n}]$.

Let us now suppose the easy-axis magnet to host a domain wall of width $\lambda=\sqrt{\bar{A}/\bar{K}} \ll L$ at $x=X$, with $\theta\approx0$ for $x\ll X$ (regions $\text{I}_{i,b}$) and  $\theta\approx\pi$ for $x\gg X$ (region III). Then, in regions $\text{I}_{i,b}$, the  exchange interaction $U_{\text{int}}\approx-g n_{z}$  leads to a tilt of the $z$ component, $n^{\text{I}}_{z}$, of the order parameter $\mathbf{n}$, with $n^{\text{I}}_{z}\gtrless0$ for $g\gtrless0$. In region III, we have instead $U_{\text{int}}\approx g n_{z}$, and the tilt, $n^{\text{III}}_{z}$, reverses its sign. Note that, for simplicity, we are taking the coupling $g$ to be weak, so that $|n^{\text{I,III}}_{z}| \ll 1$; to linear order in $g$, we neglect the tilt $\delta m_{z}$ induced on $\mathbf{m}$. $n^{\text{I,III}}_{z}$ are constant in regions $\text{I}_{i,b}$ and III, respectively, with $n^{\text{III}}_{z}=-n^{\text{I}}_{z}$.

The static canting of the magnetization, $n^{\text{I}}_{z}$,  enables the two-fluid character for the out-of-plane polarized spin transport in the easy-plane magnet.\cite{flebusPRL16} Namely, in the interfacial region $\text{I}_{i}$, the heat flux at the left interface induces a pileup of thermal magnons with chemical potential $\mu$ [see Fig.~\ref{figure2}(b)], which feed the superfluid current according to the term $\propto \eta n_z^{\text{I}}\mu$ in Eq.~(\ref{EQ12}). This gives rise to a $z$-polarized superfluid current density, which is proportional to the gradient of the azimuthal angle, i.e., $j \sim - \partial_{x} \varphi$.\cite{soninJETP78,takeiPRL14}
In region II, the coupling $U_{\text{int}}$ locks the azimuthal angles of the easy-axis and easy-plane magnets, $\phi=\varphi$ at $x=X$, impeding the superfluid current flow. Since the \textit{U}(1) symmetry demands the conservation of the $z$ component of the angular momentum, the superfluid current is absorbed by the domain wall [see Fig.~\ref{figure2}(c)] and converted into its sliding motion [see Fig.~\ref{figure2}(a)].

\subsection{Coupled dynamics}

Within the Landau-Lifshitz-Gilbert phenomenology and by including the relevant thermomagnonic torques [see Eq.~(\ref{EQ12})], the orientational order-parameter dynamics in our bilayer can be written as 
\begin{align}
\hbar(1+ \bar{\alpha} \mathbf{m} \times)  \dot{\mathbf{m}} =& -\mathbf{m} \times \delta_{\mathbf{m}} \mathcal{F} /\bar{s}\bar{t} \,, \label{eq:1} \\
\hbar(1+ \alpha \mathbf{n} \times) \dot{ \mathbf{n}} =& - \mathbf{n} \times \delta_{\mathbf{n}} \mathcal{F}/st \nonumber \\
&- \eta n_z \mathbf{n} \times (\hbar n_{z} \dot{\mathbf{n}}-\mu \mathbf{n}\times\hat{\mathbf{z}})    \,,
\label{eq:2}
\end{align}
where $s$ ($\bar{s}$) is the equilibrium spin density of the easy-plane (easy-axis) ferromagnet respectively, and the functional derivatives $\delta$ are taken with respect to the $xy$ coordinates only. (We are hereafter dropping the tilde on $\tilde{s}$.) Soft dynamics of the easy-axis ferromagnet reduces to the dynamics of the domain-wall region, which, in the collective-coordinate approach\cite{thielePRL73,*tretiakovPRL08} and using the Walker ansatz for the magnetization profile, i.e., $\ln \tan(\theta/2)=(x-X)/\lambda$, reads as
\begin{align}
\bar{s} \dot{\Phi} - \bar{\alpha} \bar{s} \dot{X}/\lambda=0\,,~~~\bar{s} \dot{X} + \bar{\alpha} \bar{s} \lambda \dot{\Phi}=\tau_{\Phi}/2\bar{t} \,.
\label{eq:13}
\end{align}
Here, the soft-mode coordinates $X$ and $\Phi\equiv\phi(X)$ are the location of the domain wall and the azimuthal angle at its center, respectively,  while $\hbar\tau_{\Phi}\equiv-\partial_{\Phi} \int dx \,U_{\text{int}}= g\int_{\lambda} dx \, \sin\theta \sin(\varphi-\phi) $  is the torque arising from the exchange interaction with the easy-plane ferromagnetic sublayer.

To linear order in $n_z$, the $z$-projected dynamics of Eq.~(\ref{eq:2}) become
\begin{align}
\hbar s (\dot{n}_{z} + \alpha  \dot{\phi}) =&A \partial^{2}_{x} \varphi - \eta s n_{z} \mu + (g/t) \sin \theta \sin (\phi - \varphi) \,.
\label{EQZ8}
\end{align}
Viewing Eq.~(\ref{EQZ8}) as a continuity equation for the $z$ component of the spin density $s_{z}=s n_{z}$ allows us to identify $j=- A \partial_{x} \varphi$ as the $z$-polarized superfluid spin-current density flowing in the $x$ direction. The thermal-magnon density $\tilde{n}$ evolves according to Eq.~\eqref{EQ13} as
\begin{align}
\dot{\tilde{n}} + \partial_{x}\tilde{j}+ \sigma \mu/\lambda_{m}^{2}=  - \eta s n_{z}\dot{\varphi}\,,
\label{eq:15}
\end{align}
where $\tilde{j}=- \sigma \partial_{x} \mu - \zeta \partial_{x} T$ (with  $\zeta$ being the bulk magnon Seebeck coefficient) is the thermal-magnon flux and $\lambda_m=\sqrt{\sigma/(\gamma+\eta s/\hbar)}$ is the thermal-magnon diffusion length, which is reduced by the superfluid coupling $\eta$.\cite{Note2} Note that so far, we are not including the direct thermomagnonic torques\cite{kovalevEPL12,kimPRB15ll} by the thermal gradient $\partial_x T$ onto the precessional order-parameter dynamics. We will comment on those below.

In the following, we solve Eqs.~(\ref{eq:13})-(\ref{eq:15}), looking for solutions of the form $\dot{\Phi}=\Omega$, $\dot{X}=v$,  $\varphi(x,t)=f(x) + \Omega t$, $\dot{n}_{z}=0$ and $\dot{\tilde{n}}=0$, corresponding to a steady-state motion of the domain wall propelled by a superfluid spin flow. We impose hard-wall boundary conditions at $x=0$ both for the superfluid and normal components of the spin current, i.e.,
\begin{align}
\partial_{x}\varphi=0\,, \; \; \; \; \sigma \partial_{x} \mu + \zeta \partial_{x} T=0 \,.
\label{eq:16}
\end{align}
Solving Eq.~(\ref{eq:15})  with the boundary conditions (\ref{eq:16})  yields
\begin{align}
\mu(x)=\mu_{0}  e^{-x/\lambda_{m}}-\left( \lambda_{m}/\lambda_{cx} \right)^{2}n_{z}^{\text{I}}  \hbar \Omega\,,
\label{EQZ11}
\end{align}
in region I. Here, $\mu_{0}= \lambda_{m} \zeta \partial_{x} T/ \sigma$ and $\lambda_{cx}=\sqrt{\hbar \sigma/ \eta s }$  is the superfluid-thermal magnon equilibration length.\cite{flebusPRL16} Integrating Eq.~(\ref{EQZ8}) in regions I and III, with $\lambda_{m},\lambda \ll X, L$, leads us to\cite{Note3}
\begin{align}
j_- &= - \eta s n^{\text{I}}_{z} \mu_{0} \lambda_{m} - \alpha s \hbar \Omega X \,, \label{EQZ12} \\
j_+ &=  \alpha s \hbar \Omega ( L- X ) \,, \label{EQZ13}
\end{align}
where $j_\mp$ are the superflow spin currents just before and after the domain wall (i.e., region II). On the other hand, the spin-current loss within the domain-wall region, $\Delta j=j_--j_+$, equals
\begin{align}
\Delta j=  \alpha s\hbar\Omega \lambda+\hbar \tau_{\Phi}/t \,. \label{EQZ14}
\end{align}
Combining Eqs.~(\ref{EQZ12})-(\ref{EQZ14})  yields
\begin{align}
 \tau_{\Phi}/t= -\eta sn^{\text{I}}_{z} \mu_0 \lambda_{m}/\hbar - s  \alpha \Omega L \, .
\label{eq:20}
\end{align}
The physical interpretation of Eq.~(\ref{eq:20}) is straightforward: The amount of the angular momentum transferred from the superfluid to the domain wall is proportional to the spin current fed into the superfluid by the thermal cloud, minus the net current loss due to Gilbert damping.
 
Using Eqs.~(\ref{eq:13}) and (\ref{eq:20}), we can determine the velocity $v$ at which the domain wall moves as
\begin{align}
v\equiv\dot{X}= - \frac{\eta n^{\text{I}}_{z} \lambda^{2}_{m} (s/\bar{s})(\zeta/\sigma)/2\hbar}{ ( \bar{t}/ t) (1+\bar{\alpha}^{2}) +   \alpha\bar{\alpha}L/2\lambda} \partial_{x} T
  \,.
\label{eq:21}
\end{align}
Equation (\ref{eq:21}) is the central result of our calculation. The numerator is proportional to the torque exerted by the thermally-induced magnon pileup at the left edge of the bilayer, while the denominator is augmented by the Gilbert-damping spin leakage associated with the domain-wall dynamics in the easy-axis layer and the superfluid dynamics in the easy-plane layer. When a ferromagnetic (antiferromagnetic) exchange coupling between the two layers is switched on, $g\gtrless0$, the superfluid induces a domain-wall motion towards the right (left) end, with a driving force proportional to the interfacial temperature gradient $\partial_{x} T<0$ and to the strength of the interaction between the superfluid and the thermal cloud within the easy-plane magnet, $\eta$.

Let us compare, in the limit of small damping, i.e., $\alpha, \bar{\alpha} \ll 1$,  Eq.~(\ref{eq:21}) with the result of Ref.~[\onlinecite{kimPRB15ll}] for the domain-wall velocity subject to the bulk thermomagnonic torques $\propto\partial_xT$. The latter concerns the domain-wall motion induced by a thermal-magnon flux traversing its profile (which should be contrasted with our superfluid-mediated torques that are induced nonlocally). Within the stochastic Landau-Lifshitz-Gilbert phenomenology, the corresponding velocity is\cite{kimPRB15ll}
\begin{align}
v\sim0.1 \frac{\partial_{x}T}{\bar{\alpha} \hbar \bar{s} \bar{\lambda}} \,,
\label{eq:22}
\end{align}
where $\bar{\lambda}=\sqrt{\bar{A}/ \bar{s}T}$ is the thermal-magnon wavelength (in units such that the Boltzmann constant is $k_B=1$). The superfluid-induced domain-wall velocity \eqref{eq:21} exceeds Eq.~\eqref{eq:22} when
\begin{align}
\eta \gtrsim \frac{0.1}{n^{\text{I}}_{z}}\frac{1}{\bar{\alpha} \bar{\lambda}} \frac{\bar{t}}{t} \frac{1}{s \lambda^{2}_{m}}  \frac{\sigma}{\zeta} \,,
\label{EQ28}
\end{align}
supposing that $\alpha\bar{\alpha}\ll(\lambda/L)(\bar{t}/t)$. Let us take, consistently with our approximations, $n_{z} \sim0.1$.
Following the transport theory of Ref.~(\onlinecite{flebusPRL16}) [Supplemental Material], we can set $\sigma / \zeta \sim 1$ in the simplest diffusive limit. Rewriting $s\sim1/a^{3}$, with $a$ being the atomic-lattice constant, Eq.~(\ref{EQZ8}) reads as
\begin{align}
\eta \gtrsim\underbrace{\left(\frac{1}{\bar{\alpha}}\right)}_{\gg1} \underbrace{\left(\frac{a^{3}}{\bar{\lambda} \lambda^{2}_{m}}\right)}_{\ll 1} \frac{\bar{t}}{t} \,.
\label{EQZ29}
\end{align}
This show that it is in principle possible to achieve domain-wall motion with the superfluid-mediated spin transfer, which is faster than the motion in response to the direct thermal gradient. We note, however, that the dissipation of energy in the superfluid case scales more favorably with the geometric dimensions of the structure, as the spin current can be supplied predominantly to the domain wall, without the diffusive/Ohmic losses throughout the entire system.

\section{Discussion and conclusion}

In this work, we have outlined a phenomenological approach to derive local thermomagnonic torques and pumping allowed by the symmetries of a magnetic system, using axially-symmetric \textit{U}(1) magnets as a concrete illustrative example.
Our formalism, which relies on the Onsager and Neumann's principles, can be extended to other classes of magnetic systems, as well as the nonlocal torque/pumping phenomena. For simplicity, we have included the dissipative spin angular-momentum losses perturbatively, disregarding their effect on the torque/pumping process. In the opposite regime of a very strong spin relaxation, this assumption can also be easily relaxed.

As a possible practical application, we have discussed the coupling between a superfluid and a domain-wall in an easy-plane|easy-axis ferromagnetic heterostructure. We have shown that a local heat flux can induce a distant motion of a domain wall via a spin superfluid. Furthermore, we have established that the transfer of angular momentum in our set-up can be more efficient than the one involving bulk temperature gradients and the direct interaction between thermal magnons and the domain wall.

Our findings allow to bridge thermal biases with collective spin dynamics, paving a way for the conversion of heat into long-ranged spin transport that suffers little dissipation. In particular, this can be used for channeling spin currents into topological soliton motion from featureless heat sources. A possible future application is the injection of chiral domain walls by means of a local thermal bias, as a natural extension of the proposal put forward in Ref.~[\onlinecite{kimPRB15br}].

\acknowledgments

The authors thank Se Kwon Kim for helpful discussions. This work is supported by FAME (an SRC STARnet center sponsored by MARCO and DARPA), the ARO under Contract No. 911NF-14-1-0016, the Stichting voor Fundamenteel Onderzoek der Materie (FOM), and the D-ITP consortium, a program of the Netherlands Organization for Scientific Research (NWO) that is funded by the Dutch Ministry of Education, Culture, and Science (OCW).

\end{document}